# Anchor Nodes Positioning for Self-localization in Wireless Sensor Networks using Belief Propagation and Evolutionary Algorithms

Saeed Ghadiri
*K.N. Toosi University of technology*
saeed.ghadiri@alumni.kntu.ac.ir

*Abstract*— Locating each node in a wireless sensor network is essential for starting the monitoring job and sending information about the area. One method that has been used in hard and inaccessible environments is randomly scattering each node in the area. In order to reduce the cost of using GPS at each node, some nodes should be equipped with GPS (anchors), Then using the belief propagation algorithm, locate other nodes. The number of anchor nodes must be reduced since they are expensive. Furthermore, the location of these nodes affects the algorithm's performance. Using multi-objective optimization, an algorithm is introduced in this paper that minimizes the estimated location error and the number of anchor nodes. According to simulation results, This algorithm proposes a set of solutions with less energy consumption and less error than similar algorithms.

## I. INTRODUCTION

Wireless sensor nodes with low energy consumption can be used to monitor an area. These small nodes communicate wirelessly using batteries as their energy source. Each node has sensors that measure the environment's temperature, pressure, light, etc. Since these nodes do not require any infrastructure, they became popular over the past years [1]. In these networks, each node performs independently but only has a limited processor, memory, and energy source. Managing the consumption of energy is crucial since batteries are the primary source of power. Data transmission between nodes is one of the most energy-consuming functions in these networks so researchers have focused on reducing energy consumption and data transfers in wireless networks.

It is crucial to accurately determine each node's position to send accurate reports about the environment. It is expensive to equip every node of a wireless sensor network with GPS, and locating each node manually in non-accessible fields is impossible. Consequently, The best method is to calculate each node's position.

In positioning methods, Other nodes are located using information about the positions of a few nodes (Anchor nodes). Positioning methods make the implementation of wireless sensor networks inexpensive and efficient [2]. Locating Anchor nodes at random can result in more positioning estimation errors.

The major contribution of this paper is finding the optimal location of anchor nodes and reducing their number without increasing the estimation of position error. We make use of the multi-objective genetic algorithm to find the position of the anchor nodes. In each generated solution by the genetic algorithm, we use the belief propagation algorithm to estimate the location of each node. The error of the location estimation of the whole network is one of the fitness objectives, and the number of anchor nodes is the other one. The results indicate that solutions found using this method are energy-efficient in different environments and have a lower total estimated error.

## II. WIRLESS SENSOR NETWORKS SELF-LOCALIZATION

By reducing the required equipment and GPS in wireless sensor networks, positioning methods aim to decrease implementation costs. In these methods, each node calculates the distance from its neighbors by measuring the power or delay of the received signal from the other nodes [3].

Several methods have been proposed to address the self-localization problem, but many do not make use of statistical formulation to solve the problem. A few techniques use the estimation of non-observed distances by multidimensional scaling [4] and multi-radiation [5]. But some methods decrease squared error and use gaussian model as uncertainty for the problem [3,6,7]. Since the noise in the localization problem is non-gaussian, these methods produce poor estimates.

Because the belief propagation algorithm is a distributed approach[8], it is suitable for the problem. Belief propagation algorithm for wireless sensor localization [9] has been an standard approach for a while. In [10] boxed bounded sampling is used to address the non-efficient sampling of the algorithm. In another research, By sending Fourier series instead of particles, the transmission data is greatly reduced [11]. [12] and [13] pruned extra edges from the graph and algorithm runs in minimum spanning tree. In [14] weighted trees are used to reduce runtime of the algorithm. In [15] boxed belief propagation is introduced to address the non-efficient sampling issue of the algorithm.

## III. POSITIONING BY BELIEF PROPAGATION IN WIRELESS SENSOR NETWORKS

Assume there are N sensor nodes scattered in a region and anchor nodes are located in predefined areas or equipped with

GPS. The 2D location of sensor $t$ is denoted by $x_t$. In each node, the distance to neighbors is measured.

$$d_{tu} = \|x_t - x_u\| + v_{tu} \quad (1)$$

Where $d_{tu}$ is the measured distance by node t from node u, $x_i$ is the real 2D location of node i, and $v_{tu}$ is the gaussian noise. It is possible that $d_{tu}$ and $d_{ut}$ may differ from each other, so the mean is calculated and synchronized between nodes t and u. The self-localization problem can be defined as a maximum a posteriori probability problem for the position of nodes $x_t$ with observations $\{d_{tu}\}$

$$p(x_1, \ldots, x_N | \{d_{tu}\}) = \prod_{(t,u)} p(d_{tu} | x_t, x_u) \prod_t P_t(x_t). \quad (2)$$

This problem can be solved with the belief propagation algorithm. For this purpose, in each sensor, a 4D distribution of its location and the neighbor's location is created (each location is 2D). Based upon the measured distance, this distribution shows the probability that the node is connected to the other neighbor.

$$P_o(x_t, x_u) = \exp\left(-\frac{\|x_t - x_u\|^2}{2R^2}\right), \quad (3)$$

where $P_o$ is the probability of two neighbors being connected and R is the communication radius. Based on this model, if two nodes get closer to each other the probability will increase. Based on the Belief Propagation algorithm, messages between nodes can be defined as

$$m_{tu}^n(x_u) = \sum_{x_t} \Psi_{tu}(x_t, x_u) \emptyset_t(x_t) \prod_{v \in N_t/u} m_{vt}^{n-1}(x_t) \quad (4)$$

$$B_t(x_t) = \emptyset_t(x_t) \prod_{v \in N_t} m_{vt}(x_t), \quad (5)$$

where $m_{tu}^n(x_u)$ is the message from node $t$ to node $u$ and $B_t(x_t)$ is the belief of $t$ about its location. In first iteration every message is set to $m_{tu}^n(x_u) = 1$ and $B_t(x_t) = P_t(x_t)$. The belief propagation algorithm continues until convergence. Then, the maximum posterior is calculated for each node, and its position is determined. The messages from anchor nodes differ from other nodes since these nodes already know their positions. A message of an anchor node is shown in Fig. 1. Regions that match with the measured distance have a higher probability.

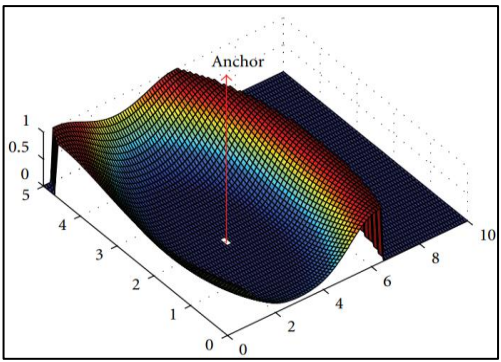

Fig. 1. Message from an anchor node[12]

The BP for localization involves first sampling from marginal distribution $\hat{P}(x_t)$ and then estimating the output message $m_{tu}$ from these samples. Each message has a 2D probability distribution of locations. The transmission of this data is expensive and heavy. Nonparametric Belief Propagation Algorithm (NBP)[9] can be used to address this issue. In NBP each message consists of probability samples. Primarily, M weighted samples from the belief probability of node t are drawn. Then using these samples and the measured distance, the message is calculated as follows

$$m_{tu}^{(i)} = x_t^{(i)} + (d_{tu} + v)[\sin(\theta^i); \cos(\theta^i)] \quad (6)$$
$$\theta^i \sim U[0, 2\pi).$$

For each sample $x_t$ drawn from the probability distribution of location, one sample is drawn randomly from a circle centered at $x_t$ and radius $d_{tu} + v$ as the message to the neighbor. These samples reconstruct the distribution in the neighbor node. Thus each sample $m_{tu}^{(i)}$ indicates a gaussian with weight $w_{tu}^{(i)}$ and covariance $\Sigma_{tu}$.

$$w_{tu}^{(i)} = P_o(m_{tu}^{(i)}) w_t^{(i)} / m_{ut}^{n-1}(x_t^{(i)}) \quad (7)$$

$$\Sigma_{tu} = M^{-\frac{1}{3}} \cdot Covar[m_{tu}^{(i)}] \quad (8)$$

All messages from neighbors are then aggregated, and then the belief of the node is built. The aggregation process of these messages and creating the belief is complicated. The Gaussian mixture method is used for weighing messages from neighbors and then drawing new samples from the distribution which is called Mixture importance sampling.[9] In Fig. 2, the nodes are assumed to be in a one-dimensional rather than a two-dimensional space. Nodes 2 and 4 are sending messages to node 3. The messages are aggregated at node 3 and the belief is created.

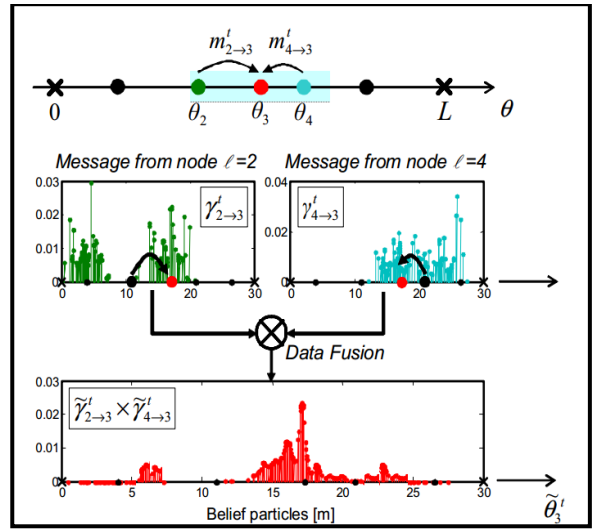

Fig. 2. Message passing in Belief Propagation algorithm in 1D space[16]

## IV. EVOLUTIONARY MULTI-OBJECTIVE OPTIMIZATION

Many real-world problems involve more than one objective. There are two objectives in the anchor node positioning problem: minimizing the estimated location error and minimizing the number of anchor nodes. It is not possible to optimize all objectives at once in multi-objective optimization[17]. So instead of one solution, a set of solutions called Pareto is needed. Pareto solutions are those that can not be better in one objective without being worse in others.

Many methods are available to solve multi-objective problems, but evolutionary algorithms are one of the most popular. These algorithms are search methods based on evolving populations. Evolution means selecting from the population and merging to reproduce new solutions[18].

Non-dominated Sorting Genetic Algorithm (NSGA 2)[19] solves the multi-objective problems with sorting solutions by the domination criteria. In this algorithm, after mutation and crossover, all solutions are sorted based on dominance and crowding distance; then, the best solutions are selected for the next generation. The algorithm steps are as follows:

1. Sort the population by dominance and create Pareto: iteratively search all non-dominated solutions that are not assigned to previous fronts in the population until all solutions have been assigned

2. Start with the first Pareto and add all of them to the next generation. Keep doing this until a Pareto has more solutions than remaining positions for the next generation.

3. Sort the last Pareto solutions with crowding distance

$$CD(d_k) = \frac{f_m(k+1) - f_m(k-1)}{f_m^{max} - f_m^{min}} \quad (9)$$

The crowding distance of the first and last solutions (border solutions) is infinite.

$$CD(d_1) = \infty, CD(d_n) = \infty \quad (10)$$

4. Fill the remaining positions with the best solutions of the last Pareto sorted by crowding distance.

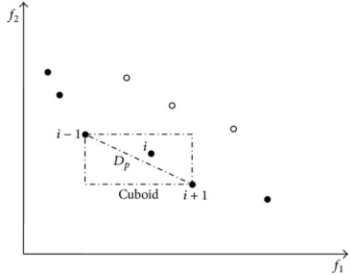

Fig. 3. Crowding Distance

With this method, the best solutions with good diversity are selected, and the algorithm uses them to generate the next population.

## V. ANCHOR NODES POSITIONING BY BELIEF PROPAGATION MULTI-OBJECTIVE OPTIMIZATION

Anchor nodes equipped with GPS are more expensive to implement, so reducing their number helps lower the implementation cost of the wireless sensor networks. Furthermore, the position of anchor nodes affects the location estimation of the entire network. Therefore, it is important to minimize their number and to place them in optimal positions. These are the two objectives that should be optimized.

To solve this problem with evolutionary methods, the chromosome's structure must be defined. Each chromosome consists of 2N genes where N is the number of anchor nodes.

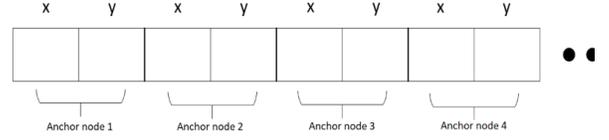

Fig. 4. Defiend Chromosome for anchor nodes positioning

As shown in Fig. 4, every 2-genes represents the 2D location of an anchor node. The population may have chromosomes with different lengths, which means that those solutions have fewer or more anchor nodes. Each chromosome's solution is given to NBP to solve the localization problem of the network. The average estimation error of the entire network is returned as a fitness function for an objective of the evolutionary algorithm. The other objective is the length of the chromosome divided by 2, which is the number of anchor nodes.

In each generation, Gaussian mutation is used to mutate the solutions. Since there are chromosomes with different lengths in the population, a customized arithmetic, one-point, and two-points crossover is used as shown in the following figures.

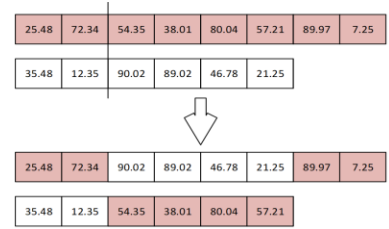

Fig. 5. Costumized on-point crossover

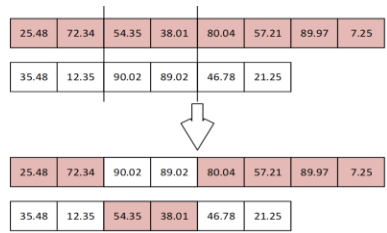

Fig. 6. Costumized two-point crossover

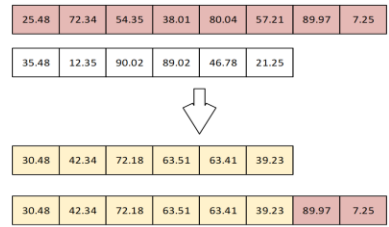

Fig. 7. Costumized arithmetic crossover

After all parents and offsprings are gathered, NSGA2 is used to select the best chromosomes for the next generation. The Algorithm runs until no improvements are made.

## VI. SIMULATION RESULTS

### A. Anchor Nodes positions and Beliefs

The simulation takes place on a square area of 100m by 100m with 100 non-anchor nodes. as shown in Fig. 8, the anchors are placed around the region in the solutions proposed by the algorithm. With these positions, the total location estimation error for the entire network is minimized.

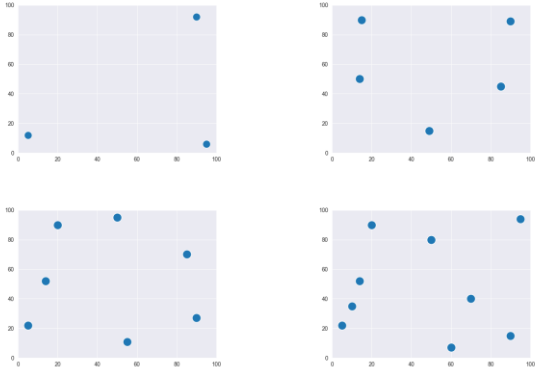

Fig. 8. Optimized position of 3,5,7, and 9 anchor nodes

The Pareto front with the number of anchor nodes and estimation error objectives is shown in the Fig. 9. It is important to note that the estimation error is increased exponentially by reducing the number of anchor nodes.

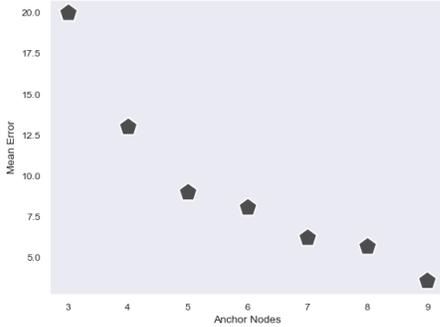

Fig. 9. Pareto solutions value for estimation error and number of anchor nodes objective

A node's belief about its 2D location over is shown in Fig. 10. The belief is converging to the real position over iterations. The estimation of location error decreases to under 4 meters after 9 iterations which is quite decent within a 100m by 100m area. This low error results in more precise reports about the area. In Fig. 11 estimation error of each node is illustrated.

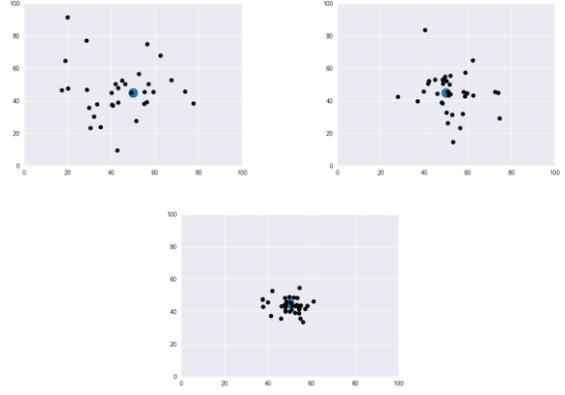

Fig. 10. Belief of a node in iterations 1,4, and 9

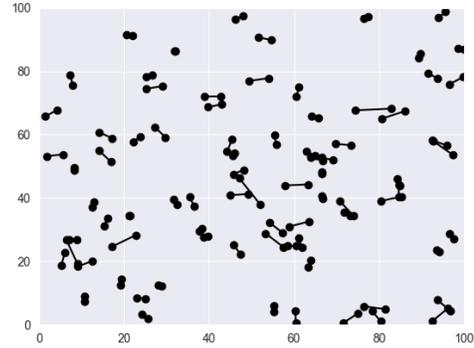

Fig. 11. Location estimation error of each node

### B. Energy consumption comparision

In this section, the energy consumption of the algorithm is compared to two similar algorithms: DV-Hop[20] and UED[21]. In DV-Hop, the average hop distance to the nearest anchor node is calculated and the node's location is estimated based on the number of hops to the anchor node. In UED, using range measurements with unknown and bounded errors the location of each node is estimated. The energy consumption of each function is defined in Table 1 in order to compare these methods energy-wise.[22]

TABLE I. ENERGY CONSUMPTION AND CONFIGURATION IN WSN

| Configuration Name | Value |
| --- | --- |
| Initial Energy | 100 J |
| Send data energy usage (per message) | 0.003 J |
| Receive data energy usage (per message) | 0.001 J |
| Run one line of code energy usage | 0.0001 J |
| Radius | 15 meter |

The experiment was repeated 10 times with nine anchor nodes. In Fig. 12, the remaining energy after using different methods is compared. There are three approaches to place anchor nodes. "MO" is the suggested multi-objective approach, "EDGE" is placing anchor nodes around the field, and "RAND" is the random placing method.

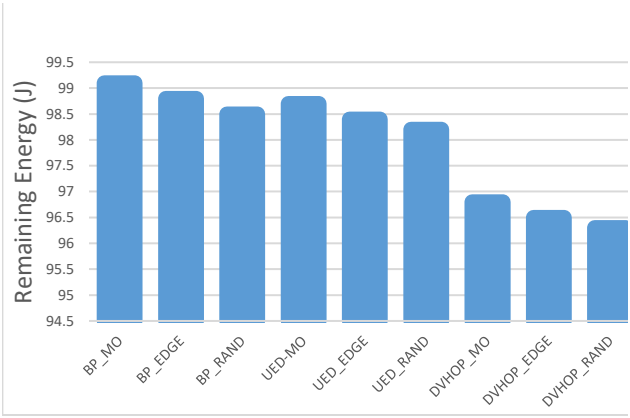

Fig. 12. Average remaining energy after localization phase for different methods

Fig. 12 illustrates that the BP-MO method consumes less energy to find the location of nodes compared to other methods. Because BP transmits messages only between direct neighbors, it uses much less energy for sending and receiving. The DVHOP method requires a lot of messages to be sent in the entire network but needs simpler calculations. you can see that random placement of the anchors results in much higher energy consumption in the network. By placing anchors near the edge of the area, energy consumption is reduced; however, the multi-objective approach is much more efficient.

TABLE II. AVERAGE ESTIMATION ERROR IN NETWORK(METER)

| Method | 3 Anchors | 6 Anchors | 9 Anchors |
|---|---|---|---|
| BP_MO | 10.511 | 8.542 | 3.956 |
| BP_EDGE | 10.024 | 7.821 | 4.786 |
| BP_RAND | 17.745 | 13.188 | 12.472 |
| UED_MO | 11.569 | 7.843 | 4.460 |
| UED_EDGE | 12.470 | 7.659 | 4.724 |
| UED_RAND | 23.633 | 17.240 | 13.070 |
| DVHOP_MO | 17.636 | 8.540 | 6.086 |
| DVHOP_EDGE | 20.031 | 8.719 | 6.764 |
| DVHOP_RAND | 22.435 | 16.757 | 14.362 |

Table 2 shows the average estimation error in the entire network with different number of anchor nodes. The results indicate that random anchor placement results in errors that are much higher than a straightforward placement near the edges. Furthermore, the multi-objective approach usually works better than other approaches.

## VII. CONCLUSIONS AND FUTURE WORKS

Wireless anchor nodes help other nodes locate themselves in the wireless networks by sending information about their location. Optimizing the location of anchor nodes not only lowers energy consumption but also results in a better location estimation of the entire network. In this paper, using multi-objective optimization and belief propagation algorithms, the optimal locations of anchor nodes were found. Implementation in 3D environment, cutting neighborhood graph for faster, reducing number of non-parametric particles in belief propagation can improve the proposed approach.